\documentclass{PoS}

\def\beq{\begin{equation}}
\def\eeq{\end{equation}}
\def\bea{\begin{eqnarray}}
\def\eea{\end{eqnarray}}
\def\beqn{\begin{eqnarray}} \def\eeqn{\end{eqnarray}}
\def\beeq{\begin{eqnarray}}
\def\eeeq{\end{eqnarray}}

\def\ep{\epsilon}

\def\nn{\nonumber}
\def\Eq#1{Eq.~(\ref{#1})}
\def\ln#1{\mathrm{log}\left(#1\right)}

\def\bq{q\hspace{-.42em}/\hspace{-.07em}}

\def\td#1{\tilde{\delta}\left(#1\right)}

\newcommand{\la}{\langle}
\newcommand{\ra}{\rangle}

\def\qb{\mathbf{q}}

\def\M#1{{\cal M}^{(#1)}}

\def\qq{{q \bar q}}

\newcommand\as{\alpha_{\mathrm{S}}}
\newcommand\g{g_{\mathrm{S}}}

\def\uv{{\rm UV}}
\def\ir{{\rm IR}}

\def\gs{g_{\rm S}}
\def\as{\alpha_{\rm S}}
\def\aas{\frac{\as}{4\pi}}
\def\v{{\rm V}}
\def\r{{\rm R}}

\def\aas{\frac{\as}{4\pi}}

\def\Se{\widetilde{S}_{\ep}}

\title{
\vspace*{-2.0cm}
\begin{minipage}{\textwidth}
{\normalfont\small IFIC/16-80
\hspace{\fill} November 2016
}\\
\end{minipage}\\[60pt]
Higher-orders in heavy quark processes within the LTD approach}

\ShortTitle{Higher-orders in heavy quark processes within the LTD approach}

\author{\speaker{Germ\'an F. R. Sborlini}$^{\ a,b}$\\\\
        $^a$Instituto de F\'{\i}sica Corpuscular, Universitat de Val\`{e}ncia -- 
Consejo Superior de Investigaciones Cient\'{\i}ficas, Parc Cient\'{\i}fic, E-46980 Paterna, Valencia, Spain.\\\
        $^b$Dipartimento di Fisica, Universit\`a di Milano and INFN Sezione di Milano,
I-20133 Milan, Italy.\\\\
        E-mail: \email{german.sborlini@ific.uv.es}}

\abstract{The computation of perturbative corrections to processes involving heavy quarks is crucial for the precision program of the LHC and future colliders. In this article, we describe a powerful approach to calculate higher-orders in QCD skipping the traditional subtraction method. Our proposal is based on the loop-tree duality (LTD) theorem, which allows to rewrite virtual contributions in terms of integrals over the real emission phase-space. Then, we proceed to combine both real and virtual contributions at the integrand level, obtaining regular expressions that can be numerically integrated in four space-time dimensions. In this way, we avoid dealing with complicated massive Feynman integrals and introducing infrared counter-terms. Some reference examples are explained, in order to exhibit the potential of our method.}

\FullConference{38th International Conference on High Energy Physics\\
                  3-10 August 2016\\
                 Chicago, USA}

\begin{document}

\section{Motivation}
\label{sec:intro}
In the last years, there has been an important revolution in the calculation of higher-order corrections to physical observables. In particular, the next-to-leading (NLO) and next-to-next-to-leading order (NNLO) corrections for multi-particle processes are a key component of this progress, since the experimental data need to be compared with the theoretical predictions at the same accuracy level. One of the most restrictive bottle-necks in these computations are related with the presence of singularities in intermediate steps, which prevents a direct numerical implementation. The usual framework consists in applying a regularization technique, such as dimensional regularization (DREG), and then remove the regularized singularities from the physical results. In order to carry out this last step, the observable under consideration must be IR-safe and a proper real-virtual combination has to be considered to fulfil the Kinoshita-Lee-Nauenberg (KLN) theorem's hypothesis \cite{Kinoshita:1962ur,Lee:1964is}. 

The subtraction method \cite{Kunszt:1992tn,Frixione:1995ms,Catani:1996jh,Catani:1996vz}, including all the recently developed variations, is based on the fact that real and virtual contributions share the same divergent structure. Thus, it is possible to define suitable counter-terms to be added to the loop amplitudes and subtracted from the real terms. However, since these contributions contain a different number of final-state particles, they are defined on different phase-spaces (PS). In particular, at NLO, the virtual contribution is associated to the Born kinematics, whilst the real-radiation term includes an additional on-shell particle in the final state. As a consequence, the counter-terms must be easily integrable to be combined with the loop amplitudes and, at the same time, they must exactly reproduce the IR singular behaviour of the real terms at integrand level. In this way, once the counter-term is subtracted from the real contribution, the remainder becomes numerically integrable.

With the purpose of overcoming the current limitations, we propose an alternative method based on the loop-tree duality (LTD) theorem \cite{Catani:2008xa,Bierenbaum:2010cy,Bierenbaum:2012th,Buchta:2014dfa,Buchta:2015wna}. The main advantage of this approach relies on rewriting the loop-amplitudes in terms of PS integrals. Thus, both real and virtual contributions become expressible in terms of the same variables, leading to a natural integrand-level combination with a fully local cancellation of singularities. This article constitutes the natural continuation of Ref. \cite{2016PROCMASSLESS} to deal with massive particles. For this reason, we will briefly describe the general lines of the four-dimensional unsubtraction (FDU) approach and will present the details that become more relevant in the massive case. In particular, the treatment of self-energies and renormalization counter-terms puts in evidence subtle differences with respect to the massless case, as well as the cancellation of quasi-collinear singularities. 

\section{Review of four-dimensional unsubtraction}
\label{sec:sec1}
As we explained in Ref. \cite{2016PROCMASSLESS}, the LTD theorem allows to express virtual amplitudes by making use of single-cuts at one-loop. These single-cuts are obtained after setting one internal line (virtual state) on-shell, modifying the prescription (with the so-called dual prescription) and replacing the loop measure by a PS integration. Using DREG to regularize the intermediate steps, the loop measure is simply $d^d q$ whilst the dual one is given by
\beq
d^d q_i \, \td{q_i} = 2 \pi \imath \, \frac{d^{d-1} \qb_i}{2 \, q_{i,0}^{(+)}} \, , \quad \quad \quad q_{i,0}^{(+)} = \sqrt{\qb_i^2+M_i^2-\imath 0}  \, ,
\label{eq:MedidaCambio}
\eeq
i.e. a typical PS-like integral. The previous discussion is straightforwardly applicable to amplitudes with single powers of the Feynman propagators. If there are higher-order poles, we must use the Cauchy's residue theorem and apply the well-known formula
\beqn
{\rm Res}({\cal A},q_{i,0}^{(+)}) &=& \frac{1}{(n-1)!} \, \left. \frac{\partial^{n-1}}{\partial^{n-1} \, 
q_{i,0}}\left({\cal A}(q_{i,0})\, (q_{i,0}-q_{i,0}^{(+)})^{n}\right)\right|_{q_{i,0}=q_{i,0}^{(+)}},
\label{eq:MultiplePole}
\eeqn
as explained in Ref. \cite{Bierenbaum:2012th}. The explicit functional dependence of the scattering amplitude affects the final form of the dual representation, due to the presence of derivatives that modify the numerator. This represents a noticeable difference in comparison with the single-pole case, where the dual representation is obtained by replacing the uncut Feynman propagators by dual propagators, and performing the sum over all the possible single-cuts.

The second crucial ingredient of the four-dimensional unsubtraction (FDU) approach is related with a momentum mapping to be applied to the real component. This step was carefully explained in Refs. \cite{Hernandez-Pinto:2015ysa,Sborlini:2016gbr,Sborlini:2016hat}, and we will briefly summarize it here. At NLO, if the Born kinematics contains $m$ particles in the final state, the real-emission term is associated with $m+1$ on-shell final state particles. Thus, we must use the Born-level momenta and the loop three-momentum $\qb$ to generate the real kinematics. To show a concrete application of the method, let's focus in the reference example of a $1 \to 2$ decay process with massive final-state particles. In that case, we start defining
\beq
p_1^{\mu} = \beta_{+}\hat{p}_1^{\mu}+\beta_{-}\hat{p}_2^{\mu}~, \quad \quad p_2^{\mu} = \beta_{-}\hat{p}_1^{\mu}+\beta_{+}\hat{p}_2^{\mu}~,
\eeq
where $\hat p_1^2=\hat p_2^2 = 0$ and $\beta_{\pm}=(1\pm \beta)/2$, with $\beta=\sqrt{1-4\,M^2/s_{12}}$. Then, we apply a partition in the real-emission PS, leading to ${\cal R}_1=\theta(y_{2r}'- y_{1r}')$ and ${\cal R}_2=\theta(y_{1r}'- y_{2r}')$ (that obviously fulfils ${\cal R}_1+{\cal R}_2\equiv 1$), where $y_{ij}'=2\, p_i'\cdot p_j'/s_{12}$. In the first region, where the radiated particle becomes collinear to $p_1'$, we propose
\beq
p_r'^\mu = q_1^\mu~, \quad\quad p_1'^\mu = (1-\alpha_1) \, \hat p_1^\mu + (1-\gamma_1) \, \hat p_2^\mu - q_1^\mu~, \quad\quad p_2'^\mu = \alpha_1 \, \hat p_1^\mu + \gamma_1 \, \hat p_2^\mu~,
\label{eq:MappingR1}
\eeq
with $q_1^2=0$, which fulfils the momentum conservation constraints by construction. In the language of the dipole-formalism, $p_2'$ represents the spectator particle, whilst the emitter is $p_1'$ and the radiated particle is $p_r'$. The spectator is used to balance momentum conservation and describe the transverse component. On the other hand, in the dual virtual amplitude, the emitter is $p_1$ and the radiated particle is represented by the cut-line $q_1$. To complete the description of the mapping, the on-shell conditions must be imposed. This leads to the equations $(p_1')^2=M^2=(p_2')^2$, which allow to compute $\alpha_1$ and $\gamma_1$. Notice that the system admits different solutions: we chose the one that is compatible with the soft-limit, i.e. $\alpha_1\to 0$ and $\gamma_1 \to 1$ for $q_1^{\mu} \to 0$. However, this is not enough to unambiguously define the transformation. So, we impose a smooth transition to the massless limit; i.e. when $M \to 0$ we recover the massless mapping available in Ref. \cite{Sborlini:2016gbr}. The advantage of this construction is that we can take the massless limit at integrand level without spoiling the numerical convergence of the integrals involved.

Finally, a completely analogous treatment has to be applied in the complementary region. These mappings can easily be extended to deal with processes with $m$ particles in the final state and different masses. Moreover, it is perfectly compatible with the massless limit by construction \cite{Sborlini:2016hat}.

\section{Renormalization at integrand level}
\label{sec:sec2}
The implementation of suitable local UV counter-terms was discussed in Ref. \cite{Sborlini:2016gbr} for the massless case. The presence of masses introduces some subtleties in the treatment of self-energies and vertex corrections. In particular, the mass acts as an IR regulator, preventing some collinear singularities to take place but still leading to soft divergences. On the other hand, since we are looking for a complete local cancellation of singularities and a smooth massless transition, it is necessary that the expressions for the massive case reduce to those already available for massless processes.

Let's start with the well-known expression for the wave-function renormalization constant\footnote{As shown in Ref. \cite{Sborlini:2016hat}, an analogous treatment can be done for the mass renormalization constant, $\Delta Z_M$.}, in the Feynman gauge with on-shell renormalization conditions, i.e.
\beq
\Delta Z_2 = \aas \, C_F \left(-\frac{1}{\ep_{\uv}}-\frac{2}{\ep_{\ir}}+ 3\,  \ln{\frac{M^2}{\mu^2}}-4\right)~,
\label{eq:SEDeltaZ2expressionINTEGRADA}
\eeq
where we kept track of the IR and UV origin of the $\ep$-poles within DREG. The unintegrated expression \cite{Sborlini:2016hat} is given by
\beqn
\Delta Z_2(p_1) &=& -\g^2 \, C_F \, \int_{\ell} G_F(q_1) \, G_F(q_3) \, \left((d-2)\frac{q_1 \cdot p_2}{p_1 \cdot p_2}+4 M^2 \left(1- \frac{q_1 \cdot p_2}{p_1 \cdot p_2}\right)G_F(q_3)\right)~,
\label{eq:SEDeltaZ2expression}
\eeqn
which includes higher-order powers of the propagators. After integration  \Eq{eq:SEDeltaZ2expression} leads to \Eq{eq:SEDeltaZ2expressionINTEGRADA} as expected. Also, notice that the corresponding formula for the massless case \cite{Sborlini:2016hat} can be recovered by simply considering $M \to 0$. The term proportional to $M^2$ is responsible of soft divergences that appears when $q_1$ is set on-shell, and it vanishes as $M \to 0$ since soft-singularities are absent in the massless self-energies. On the contrary, the collinear singularities that appear in $\Delta Z_2(M=0)$ manifest as quasi-collinear divergences, i.e. terms that behave like $\ln{M^2/\mu^2}$, as shown in \Eq{eq:SEDeltaZ2expressionINTEGRADA}. In any case, \Eq{eq:SEDeltaZ2expression} exactly matches the IR-behaviour of the squared amplitudes of the real corrections, and guarantees a local cancellation of IR and UV divergences.

Once we combine the self-energy contributions with the virtual matrix-elements, there are still UV singularities present. These have to be removed by performing an expansion around the UV propagator, i.e. $G_F(q_{\uv}) = 1/(q_{\uv}^2-\mu_{\uv}^2+\imath 0)$ with $\mu_\uv$, the renormalization scale. Besides that expansion, it is also necessary to introduce the vertex renormalization constants, expressed in unintegrated form. In the Feynman gauge, the generic expression of the vertex UV counter-term reads 
\beq
{\Gamma}^{(1)}_{A, \uv} = \gs^2 \, C_F\, 
\int_\ell \left( G_F(q_\uv)\right)^3 \, 
\left[\gamma^{\nu} \, \bq_{\uv} \, {\Gamma}^{(0)}_A \, \bq_{\uv} \, \gamma_{\nu} - 
d_{A, \uv} \, \mu_{\uv}^2 \, {\Gamma}^{(0)}_{A}\right]~,
\label{UVGenericvertex}
\eeq 
where ${\Gamma}^{(0)}_{A}$ represents the tree-level vertex. In the numerator, the term proportional to $\mu_{\uv}^2$ is sub-leading in the UV-limit. Thus, the coefficient $d_{A,\uv}$ can be adjusted in order to implement the $\overline{\rm MS}$ scheme, in which the counter-term only cancels the pole (leaving unaltered the finite piece).

Thus, the UV counter-term for the wave-function renormalization constant is given by 
\beqn
\Delta Z_2^{\uv}(p_1) &=& -(d-2)\, \g^2 \, C_F \, \, \int_{\ell} (G_F(q_\uv))^2 \, \left(1+\frac{q_\uv \cdot p_2}{p_1 \cdot p_2}\right) \left(1- G_F(q_\uv)(2\, q_\uv \cdot p_1 + \mu^2_\uv)\right)\nn \\ 
&\equiv& - \Se \, \aas \, C_F \,  \left( \frac{\mu_\uv^2}{\mu^2}\right)^{-\ep} \, \frac{1-\ep^2}{\ep} ~,
\label{eq:SEParteUV}
\eeqn
whose integrated form exactly reproduces the UV pole present in \Eq{eq:SEDeltaZ2expressionINTEGRADA}. As for the vertex constants, there are sub-leading terms proportional to $\mu_\uv^2$, which are chosen so they subtract only the UV pole part from \Eq{eq:SEDeltaZ2expressionINTEGRADA}. In consequence, we can define the UV-free wave-function renormalization constant
\beq
\Delta Z_2^{\ir} = \Delta Z_2 - \Delta Z_2^{\uv}~,
\eeq
that only contains IR singularities. To conclude this discussion, it is important to emphasize that this construction is completely general and that the sub-leading terms can be adjusted to reproduce the desired scheme dependent contributions.

\section{Production of heavy-quarks at NLO}
\label{sec:sec3}

\begin{figure}[ht]
\begin{center}
\includegraphics[width=0.45\textwidth]{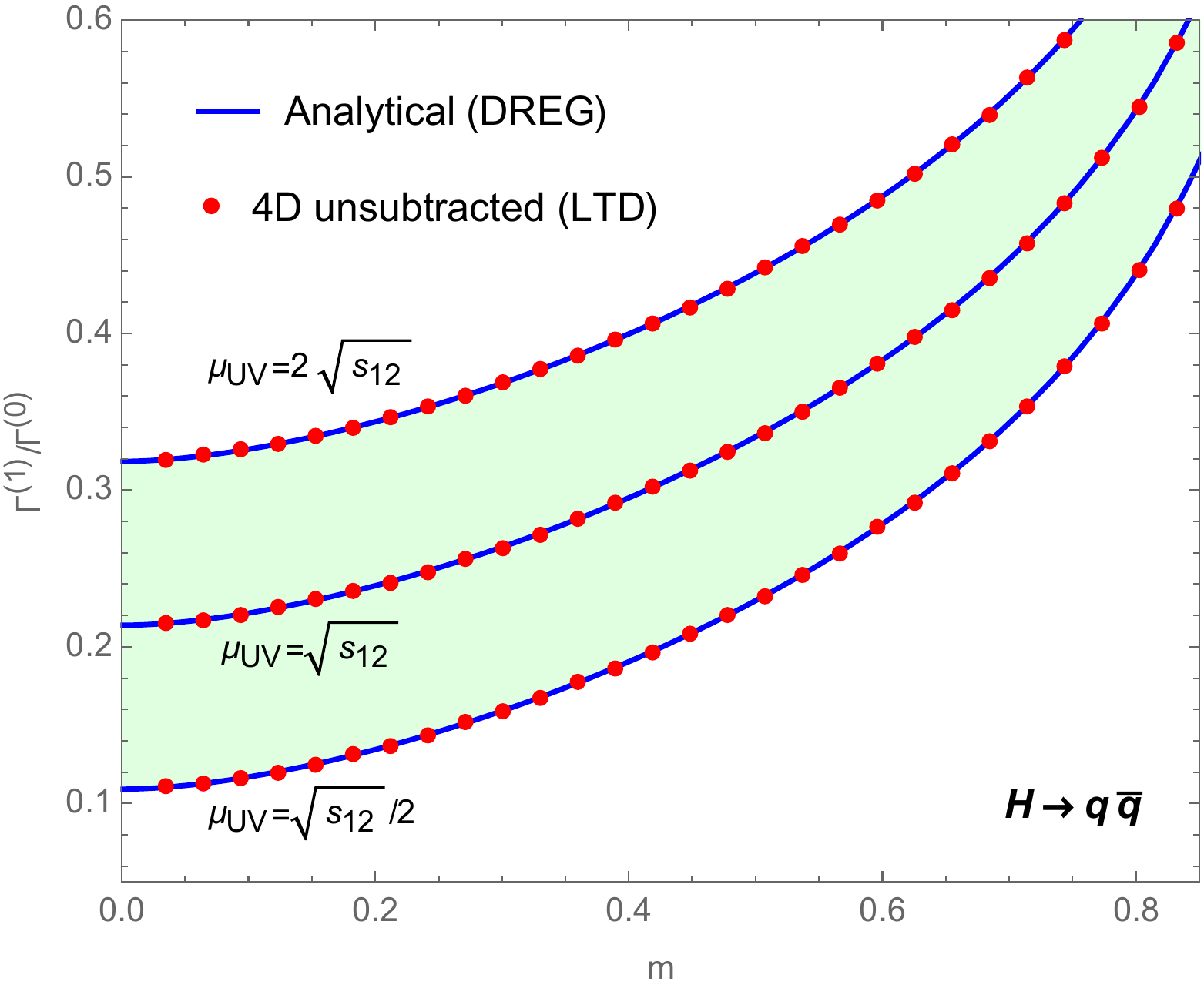} \quad
\includegraphics[width=0.46\textwidth]{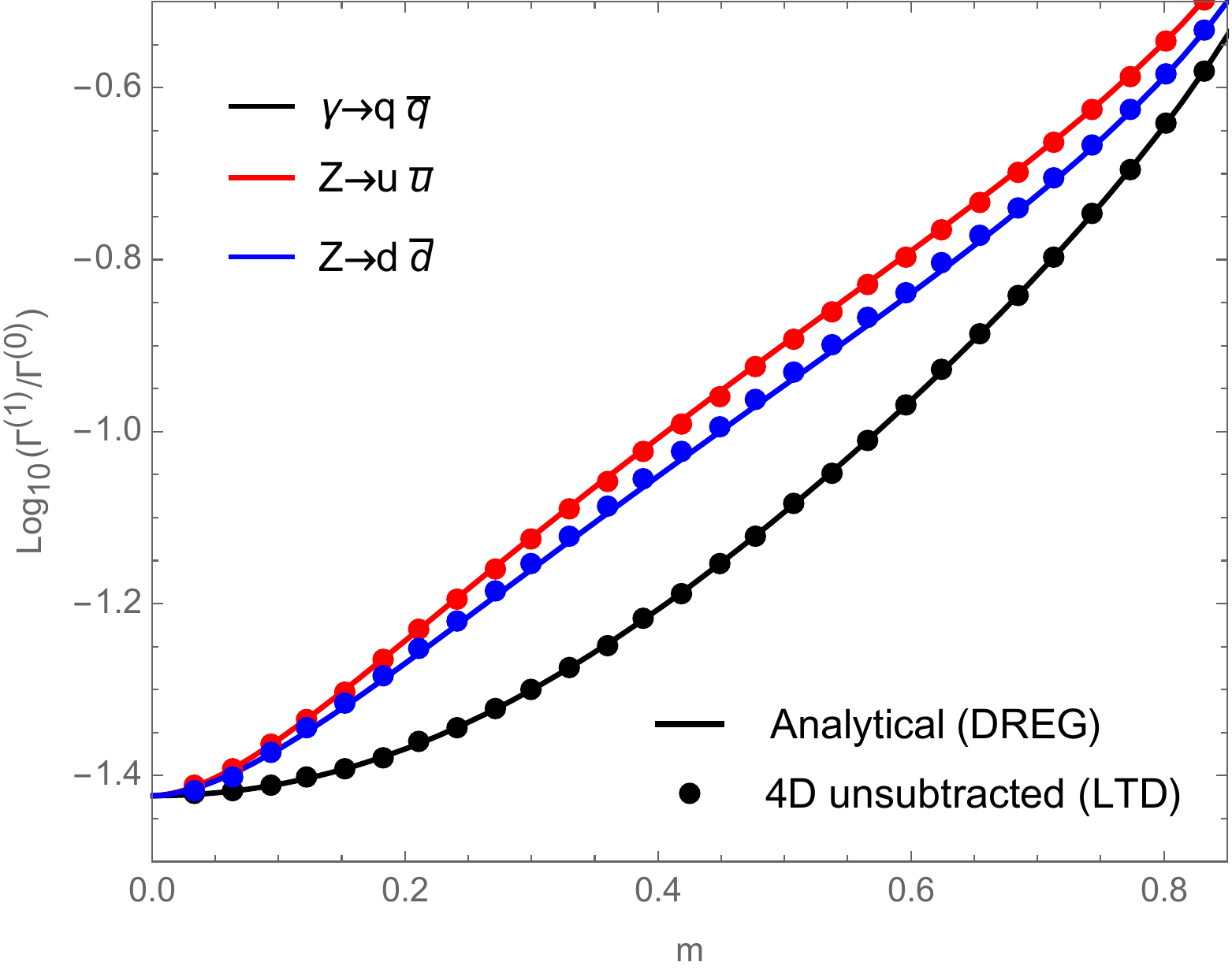}
\caption{\label{fig:Resultados}
Comparison between the analytical DREG formulae (solid lines) and the numerical FDU approach (colored dots) for different processes involving massive quarks, as a function of the dimensionless mass parameter $m=2M/\sqrt{s_{12}}$. (Left) Normalized decay rate for $H \to q \bar q$ and its scale dependence (green bands). (Right) Normalized decay rate of $\gamma$ or $Z$ boson into a massive quark-pair. In this case, there is not renormalization scale dependence because of the vector nature of the decaying particle.}
\end{center}
\end{figure}

With the ingredients described in the previous sections, we can implement the FDU approach to benchmark physical processes at NLO accuracy. For instance, we consider the NLO QCD corrections to the decay rate $A^* \to q \bar q(g)$, with $A=H, \gamma, Z$, although the algorithmic procedure is completely process-independent. Combining the renormalization factors with the one-loop amplitude, we obtain
\beq
|\M{1,\r}_A\ra =  |\M{1}_A\ra - |\M{1,\uv}_A \ra + \frac{1}{2} \left( \Delta Z_2^{\ir}(p_1) + \Delta Z_2^{\ir}(p_2) \right) |\M{0}_A\ra~, 
\label{eq:AmplitudRenormalizada1loop}
\eeq
with $|\M{1,\uv}_A \ra$ the unintegrated UV counter-term of the one-loop vertex correction, $|\M{1}_A\ra$, and $\Delta Z_2^{\ir}(p_i)$ the IR components of the quark and anti-quark self-energy corrections. Then, we apply the LTD to build the dual representation, which is given by
\beqn
\Gamma_{\v,A}^{(1,\r)} &=& \frac{1}{\sqrt{s_{12}}} \ \sum_{i=1}^3 \  
 \int d\Phi_{1\to 2} \ {\rm Re} \la {\cal M}_A^{(0)} | {\cal M}_A^{(1,\r)}(\td{q_i}) \ra~.
\label{eq:PHYSvirtualDUAL}
\eeqn
Notice that $|\M{1,\r}_A\ra$ only contains IR singularities, that must cancel those present in the real contribution by virtue of KLN theorem. The real-radiation contribution can be expressed as
\beq
\widetilde{\Gamma}^{(1)}_{\r,A,i} = \frac{1}{2 \sqrt{s_{12}}} \, \int\, d\Phi_{1\to 3} \, | \M{0}_{A\to \qq g}|^2 {\cal R}_i\left(y'_{ir} < y'_{jr}\right) \, , \quad \qquad i,j=\{1,2\}~,
\label{eq:PHYSrealDUAL}
\eeq 
where we split the integration domain and implemented the corresponding momentum mapping, as explained in Sec. \ref{sec:sec2}, and $\Gamma^{(1)}_{\r,A}=\widetilde{\Gamma}^{(1)}_{\r,A,1}+\widetilde{\Gamma}^{(1)}_{\r,A,2}$ is the real total decay rate.

The last step consists in adding \Eq{eq:PHYSvirtualDUAL} and \Eq{eq:PHYSrealDUAL}, and unifying the dual coordinate system. Expressing all the cut contributions with the same coordinate system allows to cancel divergences among the different single-cuts, thus leading to an integrable function in four-dimensions. The results are shown in Fig. \ref{fig:Resultados}, where we plot the total decay rate $H \to q \bar q$ (left) and $A^{\mu} \to q \bar q$ (right) as a function of the dimensionless mass $m=2\,M/\sqrt{s_{12}}$. As expected, a complete agreement with the analytical expressions (calculated within DREG) is obtained. Moreover, the transition to the massless limit is completely smooth, due to the properties of the momentum mapping and the definition of the local renormalization counter-terms. In addition, in the $H \to q \bar q$ computation, the local implementation of UV renormalization reproduces perfectly the renormalization scale dependence.

\section{Conclusion and outlook}
\label{sec:sec4}
In this article, we presented a brief explanation of the four-dimensional unsubtraction (FDU) approach. This technique allows to compute IR-safe observables in four space-time dimensions, without the need to introduce any regularization in the intermediate steps. Due to the application of the loop-tree duality (LTD) theorem and a physically motivated momentum-mapping, we managed to combine real and virtual contributions at the integrand level, achieving an expression that is free of singularities. Besides that, we discussed about the local implementation of the renormalization procedure, which becomes specially relevant when dealing with massive particles. We obtained integrand-level formulae for the wave-function renormalization constant, as well as for the vertex counter-term. With a well-defined algorithm, we were able to subtract the remaining UV singularities and to adjust the sub-leading pieces in order to reproduce the standard results in the $\overline{\rm MS}$ scheme. The algorithm is process-independent and we successfully applied it to compute NLO QCD corrections to the decay rate of scalar and vector particles into a massive quark pair.

As we mentioned in Ref. \cite{2016PROCMASSLESS}, this technique can be extended to deal with multi-loop multi-particle processes. The main advantage in comparison with the traditional approaches relies on the possibility of avoiding the introduction of any intermediate regularization, thus leading to a more efficient numerical implementation.

\section*{Acknowledgments}
This research project has been done in collaboration with F\'elix Driencourt-Mangin and Germ\'an Rodrigo. This work is partially supported by the Spanish Government, and EU ERDF funds (grants FPA2014-53631-C2-1-P and SEV-2014-0398) and by GV (PROMETEU II/2013/007).


\end{document}